\documentclass[pre,preprint,showpacs]{revtex4}
\usepackage{graphicx}
\begin{document}
\title{Recovering Isotropic Statistics in Turbulence Simulations:
The  Kolmogorov 4/5th-Law}
\author{Mark A. Taylor}
\affiliation{Computer and Computational Sciences Division, Los Alamos National Laboratory, Los Alamos, NM} 
\author{Susan Kurien}
\affiliation{Center for Nonlinear Studies and Theoretical Division, Los Alamos National Laboratory, Los Alamos, NM}
\author{Gregory L. Eyink}
\affiliation{Department of Mathematical Sciences, Johns Hopkins University, Baltimore, MD}
\date{\today}

\begin{abstract} 
 One of the main benchmarks in direct numerical simulations of 
 three-dimensional turbulence is the Kolmogorov 1941 prediction 
 for third-order structure functions with homogeneous and isotropic
 statistics in the infinite-Reynolds number limit. Previous DNS 
 techniques to obtain isotropic statistics have relied on 
 time-averaging structure functions in a few directions over many 
 eddy turnover times, using forcing schemes carefully constructed to
 generate isotropic data.  Motivated by recent theoretical work which
 removes isotropy requirements by spherically averaging structure
 functions over all directions, we will present results which supplement 
 long-time averaging by angle-averaging over up to 73 directions from
 a single flow snapshot.  The directions are among those natural to a
 square computational grid, and are weighted to approximate the
 spherical average. The averaging process is cheap, and for the 
 Kolmogorov 1941 4/5ths law, reasonable results can be obtained from
 a single snapshot of data.  
 This procedure may be used 
 to investigate the isotropic statistics of any quantity of interest.
\end{abstract}
\preprint{LA-UR no.}
\maketitle
\section{Introduction\label{intro}}
Both experimental and numerical studies of turbulence have attempted 
to observe the 1941 predictions of A.N. Kolmogorov \cite{K41b} for the 
statistics of isotropic, homogeneous, fully developed turbulence 
in the limit of infinite Reynolds number in an incompressible fluid. 
A main result of the theory is the so-called ``4/5-law''
\begin{eqnarray}
\langle (\delta u_L({\bf r},{\bf x}))^3\rangle &=& -\frac{4}{5} \varepsilon r 
\label{45ths_K41}\\
\delta u_L({\bf r},{\bf x}) &=& [{\bf u}({\bf x} + {\bf r}) - {\bf u}({\bf x})]\cdot \hat{\bf r}
\nonumber\\
\hat{\bf r} &=& {\bf r}/r\nonumber
\end{eqnarray}
where $\langle\cdot\rangle$ denotes ensemble averaging. The lefthand
side of Eq. \ref{45ths_K41} is the well-known third-order longitudinal 
structure function. The length scale $r$ must lie in the inertial range 
$\eta << r << L$, sufficiently far from the large scales $L$ and the 
dissipation scales given by the Kolmogorov length  $\eta$. The mean 
energy dissipation rate of the flow is given by $\varepsilon$. The 
4/5ths law is one of the few exact, non-trivial results known in the 
theory of statistical hydrodynamics. It may be reformulated in terms 
of other components of the structure function by using the 
incompressibility constraint \cite{MYII}
\begin{equation}
\langle \delta u_L(\delta u_T)^2\rangle = 
\frac{1}{6}\frac{\partial}{\partial r}(r \langle (\delta u_L)^3\rangle )
\label{incomp}
\end{equation}
where $\delta u_T$ is a velocity increment along a vector transverse 
to the separation vector ${\bf r}$. In Eq. \ref{incomp} and henceforward, 
the vector argument ${\bf r}$ is implicit. This combined with the 
Eq.\ref{45ths_K41} gives ``4/15ths law'' and the ``4/3rds law''
\begin{eqnarray}
\langle \delta u_L (\delta u_T)^2 \rangle = -\frac{4}{15}\varepsilon r
\label{415ths_K41}\\
\langle \delta u_L |\delta {\bf u}|^2\rangle = -\frac{4}{3}\varepsilon r
\label{43rds_K41}
\end{eqnarray}
where $|\delta {\bf u}|$ denotes the total magnitude of the velocity
difference across ${\bf r}$. We will refer to the three laws given by Eqs. 
(\ref{45ths_K41}), (\ref{415ths_K41}) and (\ref{43rds_K41}), and the
related theory collectively as K$41^3$. 

The K$41^3$ results have served as invaluable benchmarks for the 
empirical study of high-Reynolds number turbulence in both experiments 
and numerical simulations. Considered as exact results, they have 
allowed investigators to assess the degree to which homogeneous, 
isotropic, and high Reynolds number conditions have been attained.
Furthermore, the derivation of the K$41^3$ relations requires a
fundamental, unproven assumption, namely, that turbulent energy 
dissipation has a strictly positive limit as viscosity tends to zero. 
Hence, the validity of the K$41^3$ relations constitute an important 
test of this basic assumption. Experiments in high-Reynolds number 
turbulence performed over the past half-century do, by and large, 
support the linear scaling in $r$ of the third-order structure
functions. The convergence to the 4/5ths coefficient is quite
slow as Reynolds number increases for large scale anisotropic 
experiments \cite{SVBSCC96,BD01}, although there is empirical 
consensus that indeed this is asymptotically the correct coefficient.
Recent numerical simulations \cite{GFN02} of isotropically forced
Navier-Stokes also emphasize the slow approach to the 4/5ths law as
the Reynolds numbers are pushed as high as computational power would 
allow. A key feature of both experimental and numerical endeavors,
is the large volumes of data required -- very long time-averages, 
extending over many integral length-scales or eddy-turnover times 
are needed to obtain adequate statistics and to observe the trend 
toward K$41^3$. 

A modified version of the 4/5ths law, which does not assume isotropy 
of the flow, now exists. Nie and Tanveer \cite{NieTan99} proved 
that the 4/3rds and consequently, the 4/5ths laws can be recovered 
in homogeneous, but not necessarily isotropic flows, 
\begin{eqnarray}
\langle (\delta u_L)^3\rangle &=& 
\lim_{T\rightarrow\infty}\frac{1}{T}\int^T_0 dt \int 
\frac{d\Omega}{4\pi}\int\frac{d{\bf x}}{L^3}[\delta u_L({\bf r};{\bf x},t)]^3 
\nonumber\\
 &=& -\frac{4}{5}\varepsilon r \label{NT45}.
\end{eqnarray}
The angle integration $d \Omega$ integrates in ${\bf r}$ over the sphere 
of radius $r$. For each point ${\bf x}$ the vector increment ${\bf r}$ 
is allowed to vary over all angles and the resulting longitudinal moments
are integrated. The integration over ${\bf x}$ is over the entire flow
domain. The integration over time $t$ extends over long-times, and 
the long time average is consistent with the ensemble average of the
original K41 theory since ergodicity allows identification 
of ensemble-averages with time-averages \cite{Frisch95}. In Eq.~\ref{NT45}, 
the integration over $\Omega$ extracts the isotropic component of a
generally anisotropic flow. This is fully consistent with recent
experimental \cite{ADKLPS98,KLPS00} and numerical \cite{ABMP99,BLMT02} 
efforts to quantify anisotropic contributions by projecting the structure
function onto a particular irreducible representation, labeled by $j=0,1,\dots$, 
of the SO(3) symmetry group. The angle-averaged Eq.~\ref{NT45} correspond 
to projecting onto the $j=0$ (isotropic) sector by integration over the sphere. 
The authors of \cite{NieTan99} do not perform numerically the average over
the sphere. However, they do point out that the direction of ${\bf r}$ matters 
strongly. In their anisotropic DNS simulation at moderate Reynolds 
number, the result of taking ${\bf r}$ along a coordinate direction gave 
very poor agreement with the laws, whereas taking ${\bf r}$ along a 
body-diagonal as defined by the square-grid gave much better agreement. 

A local version of the 4/3rds law was recently derived by J. Duchon 
and R. Robert \cite{DucRob99}. Subsequently, G.L. Eyink \cite{Eyink02}
derived the corresponding version of the 4/5ths and 4/15ths laws. 
The statement is the following: Given $any$ local region $B$ of size $R$ 
of the flow, for $r<< R$, and in the limits $\nu \rightarrow 0$, next 
$r\rightarrow 0$, and finally $\delta\rightarrow 0$,
\begin{eqnarray} 
\langle (\delta u_L)^3 \rangle_{(\Omega,B)} &=& 
\lim_{\delta\rightarrow 0}\frac{1}{\delta}\int^{t+\delta}_t d\tau
\int \frac{d\Omega}{4\pi}\int_B\frac{d{\bf x}}{R^3}
[\delta u_L({\bf r};{\bf x},\tau)]^3 \nonumber\\
&=& -\frac{4}{5}\varepsilon_B r. \label{eyink45}
\end{eqnarray}
for almost every (Lebesgue) point $t$ in time, where $\varepsilon_B$ 
is the instantaneous mean energy dissipation rate over the local 
region $B$. This version of the K$41^3$ results does not require isotropy 
or homogeneity of the flow. Long-time or ensemble averages are 
also not required as in the original K41 theory \cite{Frisch95}. 
The Duchon-Robert \cite{DucRob99} and Eyink \cite{Eyink02} 
versions of K$41^3$ are truly local in space and time.

We are motivated in the present work by the existence of isotropic
statistics embedded in anisotropic data as suggested by previous work.
It is clear that both experiments and simulations face intrinsic
difficulties in achieving the high-Reynolds numbers and isotropic limit 
required by K41 theory. Both anisotropy and finite Reynolds numbers 
conspire to shorten 
the inertial range.  Experiments have achieved Reynolds numbers several orders
of magnitude higher than simulations. The indication is that at such
high Reynolds numbers, the large scale anisotropies decay faster than
the isotropic scales, allowing the latter to dominate at small scales
\cite{KurSre00}. However, while the linear scaling in $r$ of the third-order
structure function is fairly robust, the coefficient exhibits only a
slow trend toward 4/5 as indicated by the numerical work of
\cite{GFN02}. It is clear that for
anisotropic forcing, some choices of directions for the vector increment 
${\bf r}$ are more ``isotropic'' than
others \cite{NieTan99}. 

The concept of averaging over the sphere in
order to extract the isotropic component of turbulence data has existed 
for some time 
\cite{ALP99,KSreview}. Present high-Reynolds number experiments provide 
limited data -- often only a few spatial points of data acquisition,
with vector increment directions limited by the location of the probes
and the implementation of a suitable space-time surrogation (Taylor's
hypothesis). Such configurations are not suitable to spherical
averaging. Numerical data has, in principle, complete space-time
information of the flow. However, the interpolation of square grid
data over spherical shells  has been deemed too
expensive \cite{NieTan99}, or, when some such interpolation scheme is 
implemented, has not been used at sufficiently high Reynolds 
numbers as will allow for observations of the K41 type \cite{ABMP99}. 
The new angle-averaged and ``local'' laws of
\cite{NieTan99,DucRob99,Eyink02} provide us with the theoretical impetus to investigate 
and extract the isotropic component of the flow in high-Reynolds
number anisotropic turbulence. We use a novel means of taking the
average over angles which avoids the expense and effort of
interpolating the square-grid data over spherical shells.

In section \ref{sims} we discuss the numerical method and describe the
stochastic and deterministic numerical forcing schemes used in the
past and reimplemented by us.  In section \ref{avg} we present an
easily implemented scheme to average over a finite number of
angles. Using the second and third-order isotropy relations, we
demonstrate that this scheme is a good approximation to the true
spherical average.  We then present results for the angle-averaged
third-order structure functions computed both from a snapshot of the
flow at a single instant in time, and from time-averaging the data
from several snapshots.   A summary and concluding discussion of the
results is given in section \ref{disc}.
 
\section{Numerical Simulations\label{sims}} 
The numerical simulation of the forced Navier-Stokes equation for an 
incompressible flow are given by
\begin{eqnarray}
\partial_t {\bf u} + {{\bf \omega} \times {\bf u}} + \nabla \phi &=& \nu \nabla^2 
{\bf u} + {\bf f} \\
\nabla \cdot {\bf u} &=& 0
\end{eqnarray}
where the vorticity $\omega = \nabla \times {\bf u}$ and $\phi$ is
determined so as to maintain $\nabla \cdot {\bf u} = 0$. The domain is a
periodic box of side $L=2 \pi$ with $N=512$ grid points to a side.  A standard
Fourier pseudo-spectral method is used for the spatial discretization
and the equations are integrated in time using a fourth-order
Runge-Kutta scheme.  Aliasing errors from the nonlinear term 
are effectively controlled by removing all coefficients 
with wave-number magnitude greater than $k_{max}=\frac{\sqrt 2}3 N$. The 
code is optimized for distributed memory parallel computers and uses
MPI for inter-process communication. The runs were made using 256 processors
of a Compaq ES45 cluster.  

We make use of two different types of low wave number forcing.  The
first is modeled after the deterministic forcing schemes described in
\cite{Kerr94,SVBSCC96,OvePop98}, where the energy in a few low wave
numbers is relaxed back to a target spectrum.  We refer to the output
using this forcing as the deterministic dataset.  The second forcing
is the stochastic forcing used in \cite{GFN02}, where the Fourier
coefficients of ${\bf f}$ are chosen randomly, and we refer to the
data produces with this forcing as the stochastic dataset.  Both
forcings have advantages and disadvantages.  The deterministic forcing
equilibrates quickly and has less variance in time so that less data
is needed for converged time averages.  But there is an unavoidable
anisotropy throughout the simulation if the forcing is restricted to
the lowest wave numbers.  The stochastic forcing has a larger variance
in time so that data from more snapshots is needed to obtain converged
time averages, but statistics from those snapshots are observed to be
more isotropic.  We perform both kinds of forcing in order to
demonstrate the equivalence of the results when angle-averaging is
applied to the data.

Parameters of interest for both simulations are given in Table
\ref{data_parms}.  For the stochastic forcing, we have chosen
parameters similar to those used for the $512^3$ simulations in
\cite{GFN02}.   The parameters for the deterministic
case were chosen so that $R_\lambda$ would be similar in both cases.

\begin{table}
\begin{tabular}{|c|c|c|c|c|c|c|c|}
\hline
 Data set & $N$ & $\nu$ & $\varepsilon$ & $k_{max} \eta$ & $R_\lambda$ & 
$n_s$ & $n_r$ \\\hline\hline
 Stochastic & 512 & $6\times 10^{-4}$& .5 & 1.1 & 263 & 6 & 6 \\\hline
 Deterministic & 512& $4\times 10^{-3}$& 156 & 1.1 & 249 & 1 & 6 \\\hline
\end{tabular}
\caption{The parameter values for the two data sets. N = Number of grid points
per coordinate direction; $\nu$ = viscosity; 
$\varepsilon$ = Mean energy dissipation rate 
$\eta$ = Kolmogorov scale; 
$R_\lambda$ = Taylor microscale Reynolds number; 
$n_s$ = number of eddy turnover times to spin up; 
$n_r$ = number of eddy turnover times computed after spin up. }
\label{data_parms}
\end{table}

\subsection{Deterministic forcing \label{forcing_det}}
We first define the energy in each spherical wave number $k$ in
the usual way:
\begin{equation}
E(k) = \sum_{k-.5 \le |{\bf k}| <k+.5} \frac12 | \tilde {\bf u}_{\bf k} | ^2
\end{equation}
where $\tilde {\bf u}_{\bf k}$ is the $k$th Fourier coefficient of
${\bf u}$.  We then choose a target spectrum function given by $F(k)$, which
we set to $F(1) = F(2) = .5$ and $F(k) = 0$ for $k > 2$. We generate
a velocity field $\tilde {\bf v}_{\bf k}$ with energy $F(k)$ 
by setting
\[
  \tilde {\bf v}_{\bf k} = \sqrt{\frac{F(k)}{E(k)}} \, \tilde {\bf u}_{\bf k}
\]
The Fourier coefficients of the forcing function are then
given by
\[
\tilde {\bf f}_{\bf k} = \left\{ 
\begin{array}{ll}
\tau \left( \tilde {\bf v}_{\bf k} -  \tilde {\bf u}_{\bf k} \right) , 
& \qquad F(k) > E(k)\\
0, & \qquad F(k) \le E(k)\nonumber
\end{array}
\right.
\]
The relaxation parameter is chosen as 
$\tau^{-1} = 2|  \nabla u |$, a simplified version of the formula given in 
\cite{OvePop98}.  

This forcing simply relaxes the amplitudes of the Fourier coefficients
in the first two wave numbers so that the energy matches the target
spectrum $F(k)$ in those wave numbers. It has no effect on the phase
of the coefficients.  The phases are observed to change very slowly,
giving rise to persistent anisotropy in the large scales.


\subsection{Stochastic forcing \label{forcings_sto}}

Our second type of forcing is modeled after the stochastic scheme of
\cite{GFN02} in which the wave-numbers $|{\bf k}| \le 2.5$ are
stochastically forced.  This ensures that the phase of each forced
mode changes sufficiently rapidly so that the large scales will be
statistically isotropic.  At the beginning of each time-step, we choose a
divergence-free forcing function,
\begin{equation}
{\bf f } = \nabla \times \bigtriangleup^{-1} {\bf g}
\end{equation}
where the Fourier coefficients of ${\bf g}$, denoted by 
$\tilde {\bf g}_{\bf k}$,  are chosen randomly
with uniformly distributed phase and Gaussian distributed
amplitude.  The variance in the Gaussian distribution is chosen so that
\begin{equation}
\sum_{.5 \le |{\bf k}| <1.5} | \tilde {\bf g}_{\bf k} | ^2
= \sum_{1.5 \le |{\bf k}| <2.5} | \tilde {\bf g}_{\bf k} | ^2
= 18
\end{equation}


\section{Angle-Averaging Technique \label{avg}}
We would like to extract the isotropic component of a flow by a
suitable average of the two-point structure functions over the solid
angle $\Omega$ as defined by Eqs. (\ref{NT45}--\ref{eyink45}). 
We approximate the spherically averaged third-order longitudinal
structure function by the following average over a finite number $N_d$ 
of directions:
\begin{equation}
\langle (\delta u_L(r))^3\rangle =  \frac{1}{N_d}\frac{1}{N^3}\sum_{j=1}^{N_d}
\sum_{i=1}^{N^3}
w_j\Big[ \delta u_L({\bf r}_j; {\bf x}_i)\Big]^3.
\end{equation}
where ${\bf x}_i$ denotes grid-points, ${\bf r}_j$ denotes the
increment vector in the $j$th direction, $r = |{\bf r}_j|$ is fixed,
and the $w_j$ are quadrature weights.  Here we are using 
the longitudinal structure function as an example.  The
procedure applies equally well to any two-point structure function.  

The simulation is computed on a fixed uniform rectangular mesh. Thus
we are faced with the difficulty of evaluating ${\bf u}$ at points 
$({\bf x}_i+{\bf r}_j)$, most of which will not be grid points. 
The most straightforward approach would be to perform 3-dimensional
interpolations at each of the points $({\bf x}_i+{\bf r}_j)$.  This
requires $N^3 N_d$ 3-D interpolations of the velocity-vector field for
each separation distance $r$, which is prohibitively expensive
\cite{NieTan99}.

\begin{figure}
\centering
\includegraphics[scale=0.75]{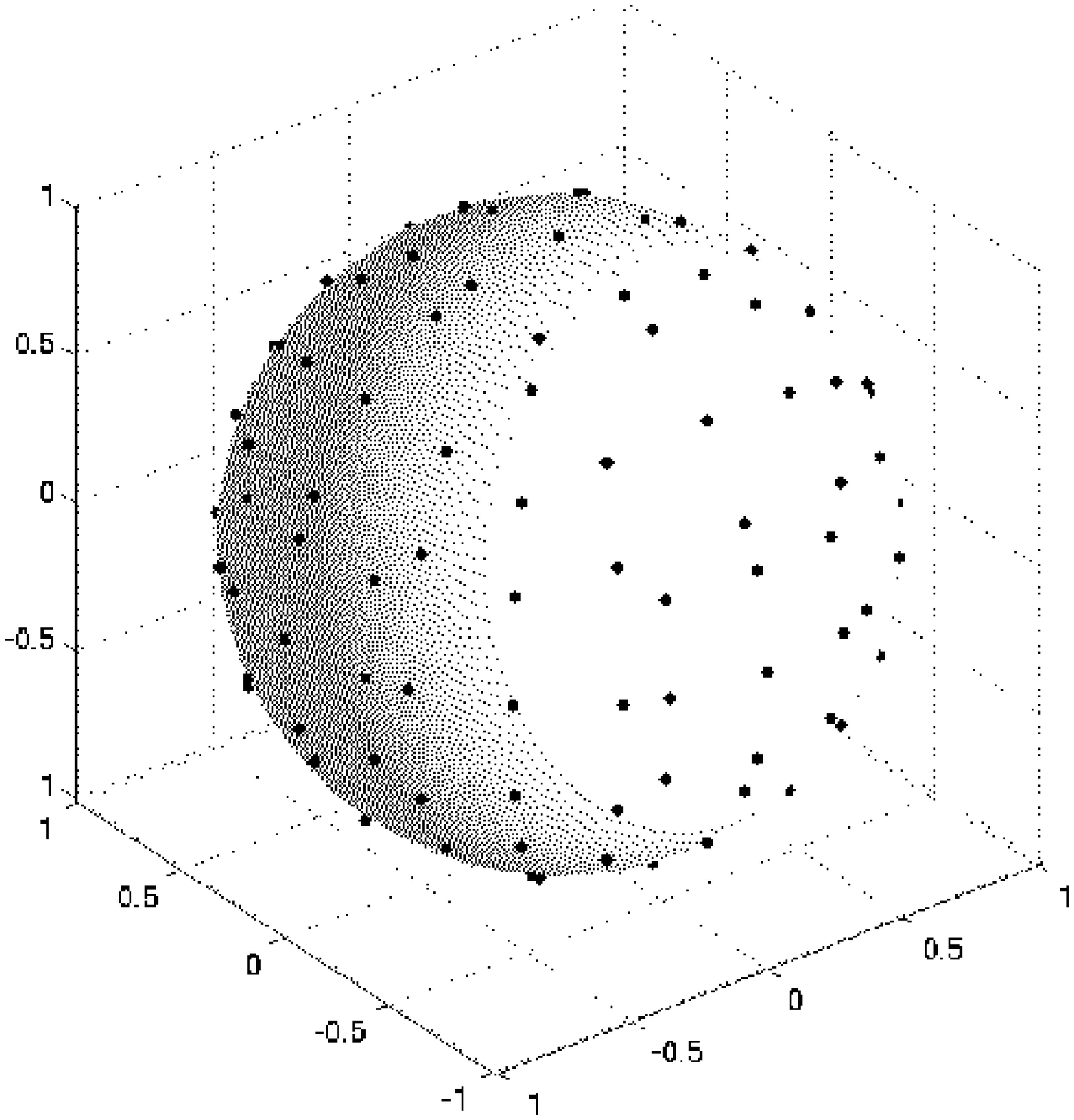}
\caption{Unit sphere showing some of the $N_d=73$ directions over which the 
average is taken. }
\label{sphere_dir}
\end{figure}

We have developed a less expensive technique for angle-averaging which
does not require any 3-D interpolations.  We first choose the vectors
${\bf r}_j$ from among those natural to a square computational
grid. We restrict ourselves to the set of all unique
directions which can be expressed with integer components with length
less than or equal to $\sqrt{11}$. 
Let $j=1 \cdots N_d$ be the index for this set. 
Each ${\bf
r}_j$ is the minimum grid-point separation distance in the
$j$th direction.  
This set is generated by the
vectors (1,0,0), (1,1,0), (1,1,1), (2,1,0), (2,1,1), (2,2,1), (3,1,0)
and (3,1,1) by taking all index and sign permutations of the three
coordinates, and removing any vector which is a positive or negative
multiple of any other vector in the set.  This procedure generates a
total of $N_d = 73$ unique directions.  The unit vectors associated
with each direction are plotted as points on the sphere in
Fig. \ref{sphere_dir}.  One can see that these points are well
distributed over the sphere.  Both the unit vectors $\hat {\bf r}_j$
and $-\hat {\bf r}_j$ are plotted, but below we do not consider the
$-\hat {\bf r}_j$ directions since they give the same contribution as
$\hat{\bf r}_j$ when averaged over the periodic computational domain.

\begin{figure}
\centering
\includegraphics[scale=1.0]{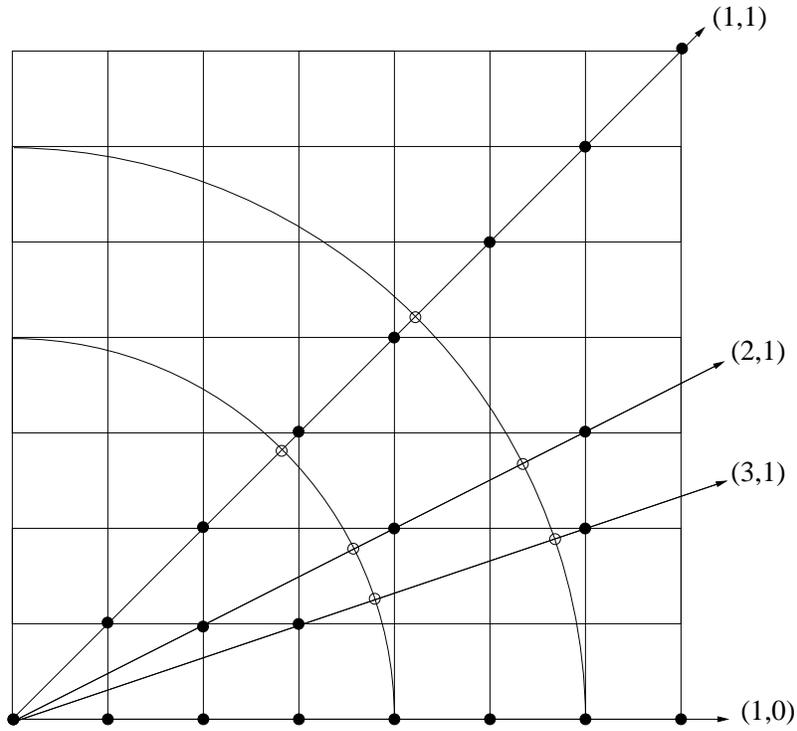}
\caption{Two dimensional example of how data is collected for the angle-averaging
procedure.  Four directions are shown, (1,0), (2,1), (3,1) and (1,1).  Velocity
data is known at all the grid points.  The black
dots represent values of $r$ where structure functions for a particular
direction can be computed with no interpolations.  Each structure function 
can then be interpolated to specific values of $r$, shown by the white dots.}
\label{grid}
\end{figure}

\begin{figure}
\centering
\includegraphics[scale=.85]{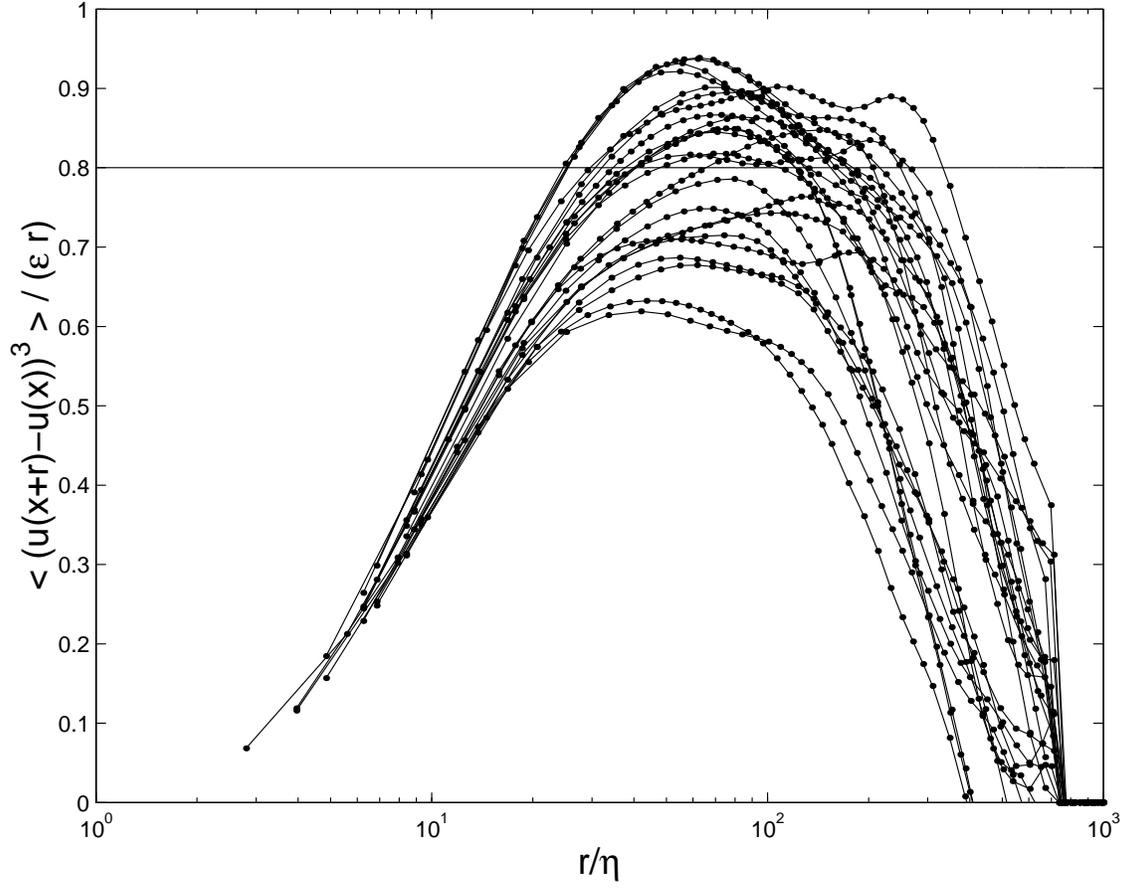}
\caption{The third-order structure function normalized by $\varepsilon r$, 
computed from a single snapshot from the deterministic dataset.
The dots indicate the values
of the structure function computed at various $\ell{\bf r}_j$.  Each thin
curve is the cubic-spline interpolation through all computed values of
the structure function in a particular direction. Only a few of the 73 
different directions are shown here for visual clarity. 
The horizontal line indicates the 4/5ths mark.}
\label{directions}
\end{figure}

For each of the $N_d$ directions, we form a set of $\ell=1 \cdots N_r$
separation vectors, ${\bf x}_i + \ell {\bf r}_j$. Since ${\bf r}_j$ is
the minimum separation distance of grid-points in the $j$th direction
and $\ell$  is an integer, all the ${\bf x}_i + \ell {\bf r}_j$ lie on 
our computational grid.  This is illustrated (in two-dimensions) for 
four directions in Fig. \ref{grid}, where the black dots represent the 
points ${\bf x}_i + \ell {\bf r}_j$ and ${\bf x}_i$ is shown at the origin.  
We can now
efficiently compute structure functions in $N_d$ different directions,
at $N_r$ separation distances for each direction, 
without any 3-d interpolations:
\begin{equation}
\langle (\delta u_L(\ell {\bf r}_j))^3\rangle =  \frac{1}{N^3}
\sum_{i=1}^{N^3}
\Big[ \delta u_L(\ell {\bf r}_j; {\bf x}_i)\Big]^3.
\end{equation}
For each direction, we get a 1-dimensional curve as a function of
$\ell{\bf r}_j$, as shown in Fig. \ref{directions}.

In the figure, points represent structure function values at the 
separation distances $ \ell |{\bf r}_j | $, and each line is a
cubic-spline fit to the data at $\ell =1 \cdots N_r$ along each 
of the $N_d$ directions. One can see that only a few directions are 
computed at each of the separation distances, so we cannot directly take 
an angle-average from this data.  But one can also see that the curves 
are quite smooth and the cubic-spline is an excellent interpolant. 
Thus we use cubic-spline interpolation to calculate the structure 
function in each of the $N_d$ directions, at separation vector 
$r \hat {\bf r}_j$ of any desired length $r$.   

Once the data for each direction has been interpolated to a common
separation distance $r$, we can approximate the angle-average at $r$
by quadrature over the $N_d$ directions:
\begin{equation}
\langle (\delta u_L(r))^3\rangle =  \frac{1}{N_d}
\sum_{j=1}^{N_d}w_j\langle (\delta u_L( r \hat {\bf r}_j))^3\rangle
\label{angave}
\end{equation}  
In order to determine the quadrature weights $w_j$, we use the software package
Stripack \cite{STRIPACK} to compute the Voronoi tiling generated by
the points ${\hat {\bf r}}_j$ on the unit sphere
centered at ${\bf x}$.  The weight $w_j$ is the solid angle subtended
by the Voronoi cell containing the point ${\hat {\bf r}}_j$. 

The angle-averaging procedure described can be implemented efficiently
on parallel computers, requiring only the same type of parallel data
transpose operator already used by a parallel pseudo-spectral code.
The total cost of this angle-averaging procedure for one snapshot
(73 directions and 100 different separation distances) 
is about the same as 150 timesteps of the Navier-Stokes code.
Thus for a single eddy turnover time, where thousands of timesteps are
required, the angle-averaging statistics can be computed during the
computation with minimal impact on the total CPU time requirement.

\subsection{Extracting the isotropic component} 

We first present results demonstrating how well the angle-average
procedure performs at extracting the isotropic component from
our DNS data.  We again follow \cite{GFN02} and examine
the relations between the second
and third order velocity structure functions:
\begin{eqnarray}
\langle (\delta u_T)^2 \rangle = (1+\frac{r}2 \frac{d}{dr} ) \langle (\delta u_L)^2 \rangle 
\label{iso2nd} \\
\langle \delta u_L (\delta u_T)^2 \rangle = \frac16 \frac{d}{dr}r 
\langle (\delta u_L)^3 \rangle  \label{iso3rd}
\end{eqnarray}
These equations require only isotropy and incompressibility.  Thus in
DNS data, where incompressibility is obtained to numerical round off
error, deviations in the above relations are a measure of the
anisotropy in the data.  In \cite{GFN02}, the left and right sides of
these equations are plotted after averaging in time.  Excellent
agreement is obtained in the inertial range, with some departure at
larger scales. 

\begin{figure}
\centering
\includegraphics[scale=.65]{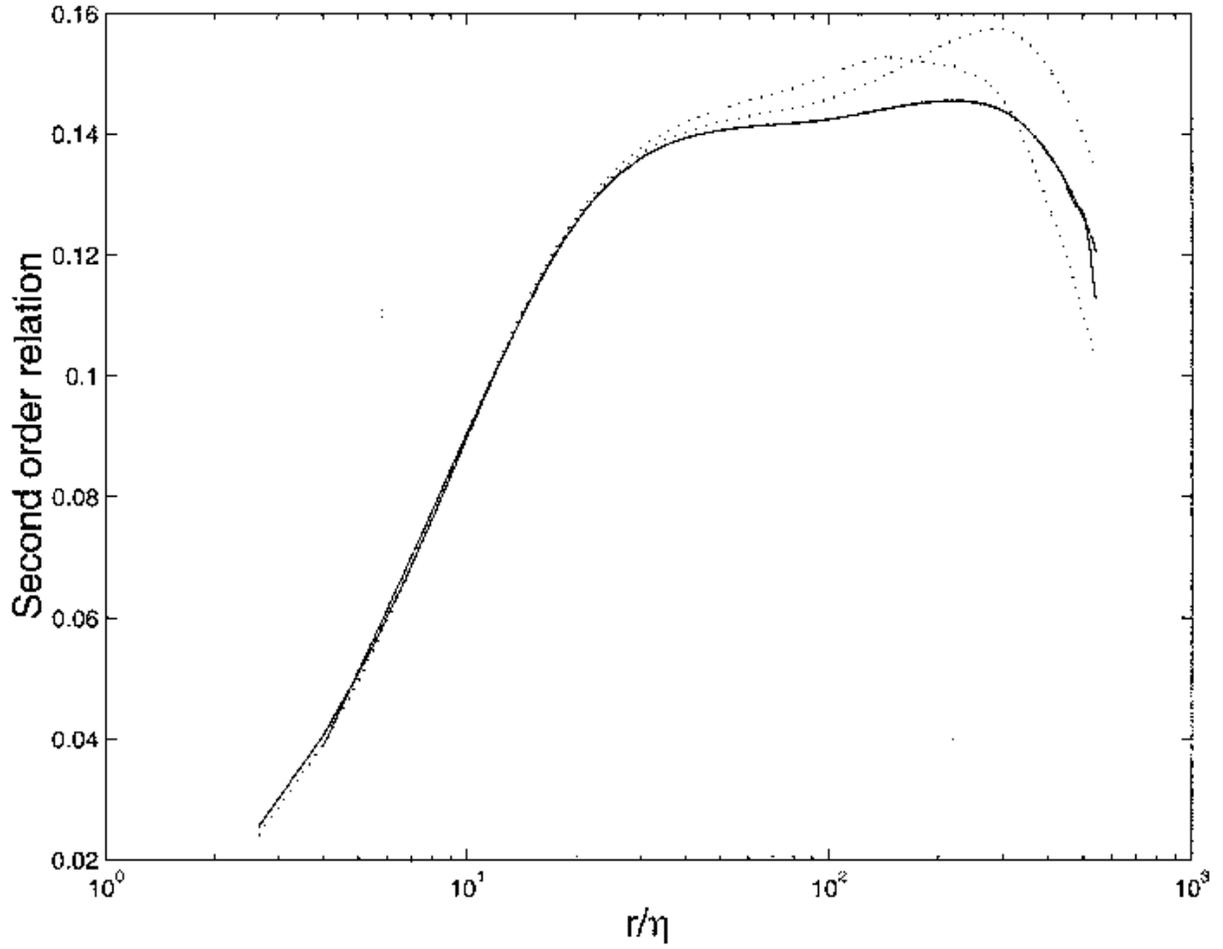}
\caption{Second order isotropy relation for the stochastic dataset. Solid lines:
left and right side of equation \ref{iso2nd}, normalized by $r^{2/3}$
and angle-averaged.  Dotted lines: same quantities, only for a single
coordinate direction.  }
\label{isocheck2}
\end{figure}

\begin{figure}
\centering
\includegraphics[scale=.85]{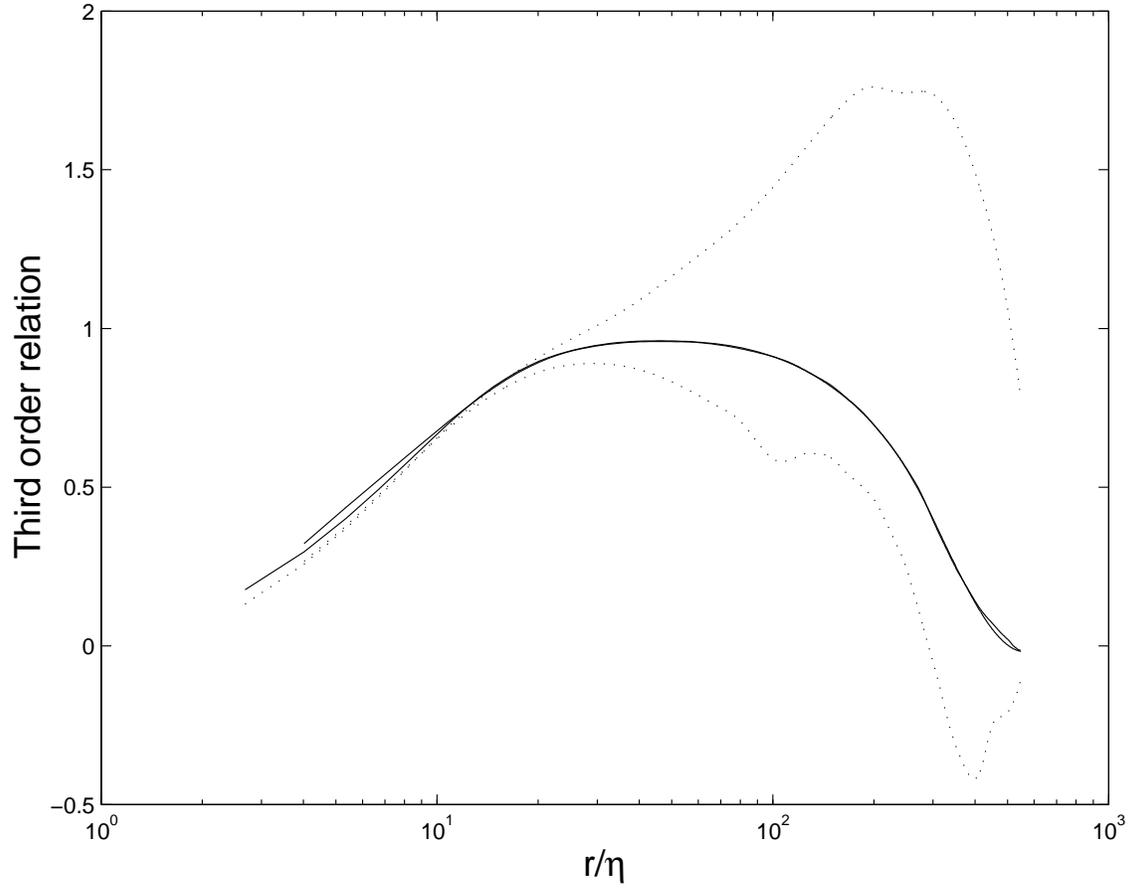}
\caption{Third order isotropy relation for the deterministic dataset.  Solid lines:
left and right side of equation \ref{iso3rd}, normalized by $r$ and 
angle-averaged.  Dotted lines: same quantities only 
for a single coordinate direction.
}
\label{isocheck3}
\end{figure}

In Fig. \ref{isocheck2}, we show the second order isotropy relation
for our stochastic dataset, and in Fig. \ref{isocheck3} we show the
third order relation for the deterministic dataset.  This data is
computed by angle-averaging over a single snapshot of the flow.  The
agreement is excellent, both in the inertial range and at the largest
scales.  For comparison, the figures also show the same relations from
the same snapshot but using only a single coordinate direction instead
of angle-averaging.  In that case, there are significant differences
for scales well into the inertial range.  Thus the angle-averaging
technique appears to be extremely effective in extracting the
isotropic component of anisotropic data, even at large scales where 
anisotropy remains after time averaging over many snapshots. Similar 
results were obtained for the second order isotropy relation from the 
deterministic dataset and for the third order isotropy relation from the 
stochastic dataset.

\subsection{Angle-averaging a single snapshot \label{aa_results}}
\begin{figure}
\centering
\includegraphics[scale=.6]{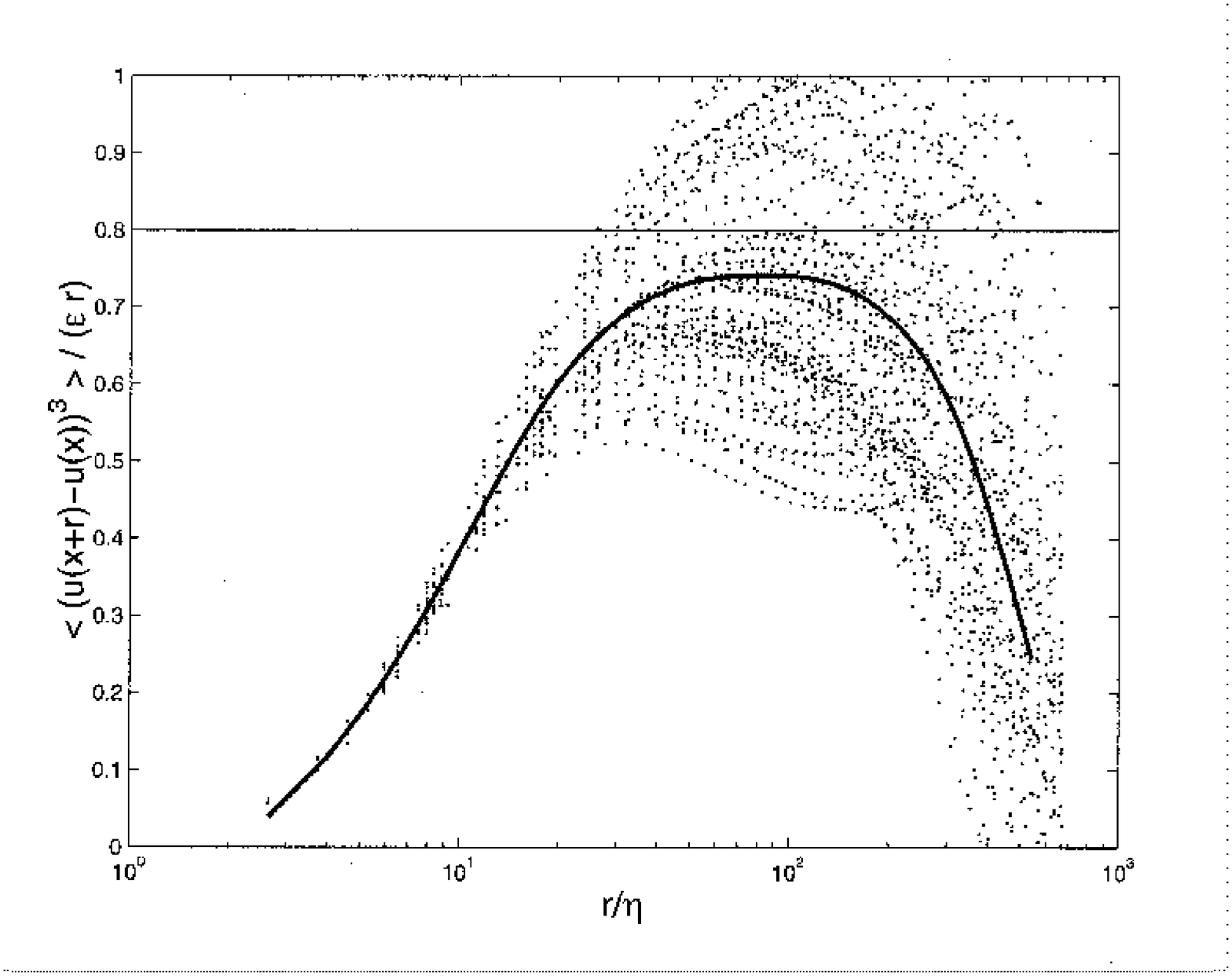}
\caption{The third-order longitudinal structure function normalized by $\varepsilon r$, 
computed from a single snapshot of the stochastic dataset.
The dots indicate the values
of the structure function computed at various $\ell{\bf r}_j$. 
The thick curve is the angle-average. The horizontal line indicates the 4/5ths mark.}
\label{sto_snapshot}
\end{figure}

\begin{figure}
\centering
\includegraphics[scale=.85]{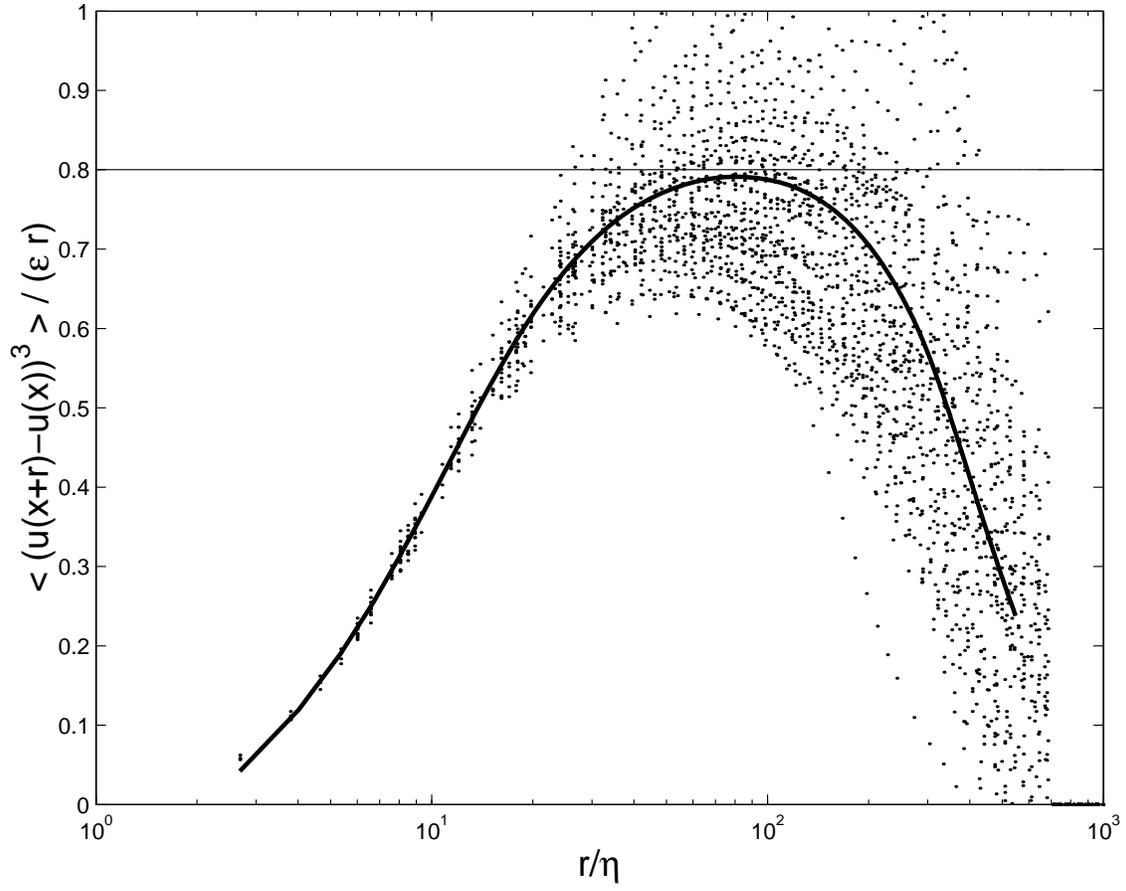}
\caption{The third-order longitudinal structure function normalized 
by $\varepsilon r$, computed from a single snapshot of the deterministic dataset.
The various symbols
and lines mean the same as in Fig. \ref{sto_snapshot}.}
\label{det_snapshot}
\end{figure}

We now present results from using angle-averaging to compute the
third-order longitudinal structure function in the 4/5ths law.  
Figures \ref{sto_snapshot} and \ref{det_snapshot} show the result of the
angle-averaging procedure described above for single snapshots of the
stochastic and deterministic datasets respectively. The snapshots are
taken after the flow has had time to equilibrate.  The value of the
mean energy dissipation rate $\varepsilon$ was calculated from the
snapshot.  This is to be contrasted with previous works in which
$\varepsilon$ is a long-time or ensemble average. We have therefore
computed a version of the 4/5ths relation which is local in time.  The
dots represent the data from all 73 directions at all values of $r$
that were computed.  The final weighted angle-average of
Eq. (\ref{angave}) is given by the thick curves in both
Figs. \ref{sto_snapshot} and \ref{det_snapshot}.  One can see that the
results from different directions are quite different, while the
angle-averaged results are quite reasonable and similar to each other
as well as similar to the results obtained from long time averaging of
the coordinate directions presented in \cite{GFN02} and shown for our
data in section \ref{tresults}.  Thus we conclude that
angle-averaging the data from a single snapshot yields a very
reasonable result.  
Similar results (not plotted) are obtained for the 4/3'rds and 4/15'ths laws.

\subsection{Temporal variance \label{timeresults}}

\begin{figure}
\centering
\includegraphics[scale=.85]{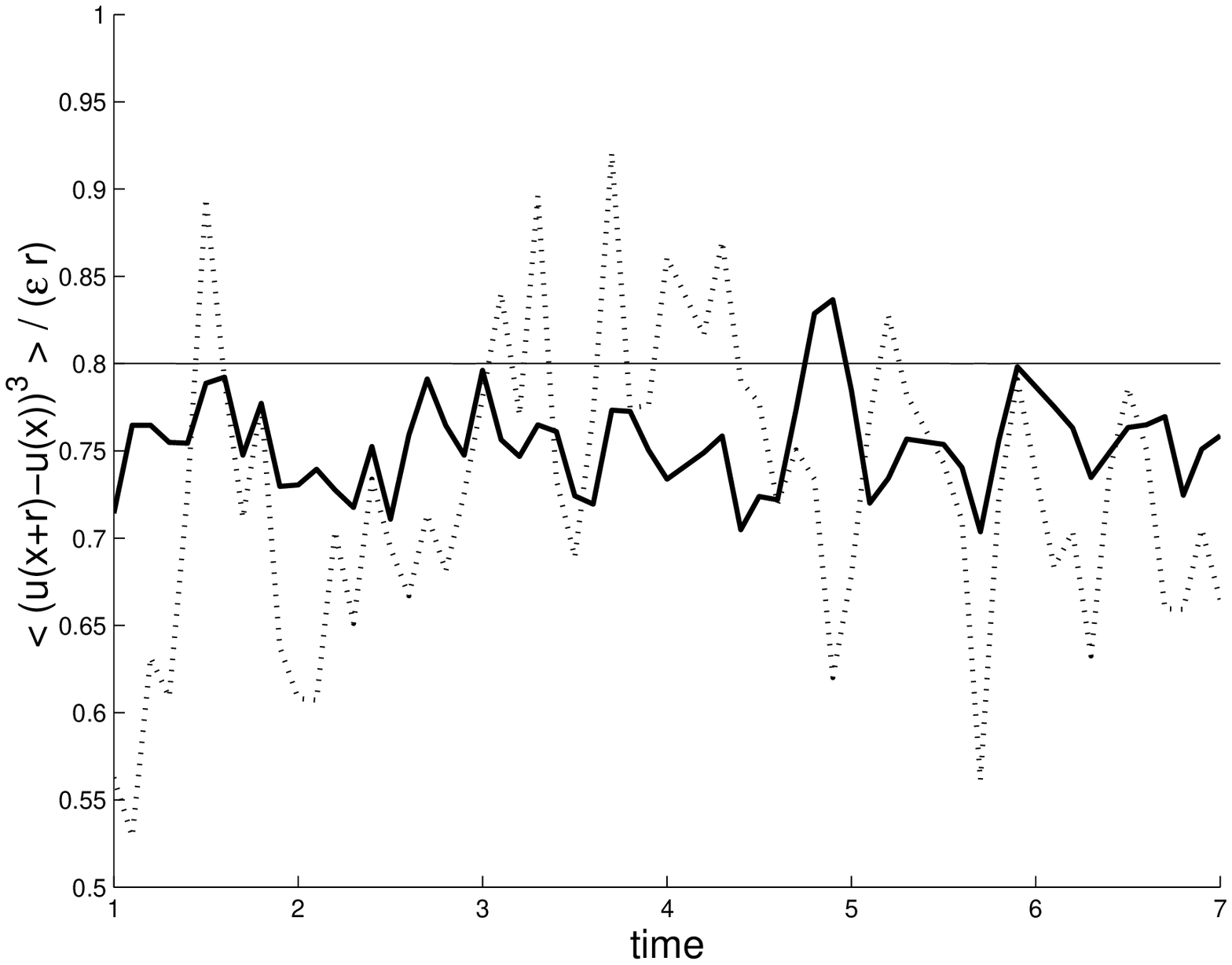}
\caption{The angle-averaged (solid line) and single-direction (dotted line) 
values of the peak of the non-dimensionalized third-order longitudinal
structure function for deterministic dataset, as a function of time.}
\label{timeseries}
\end{figure}

\begin{figure}
\centering
\includegraphics[scale=.6]{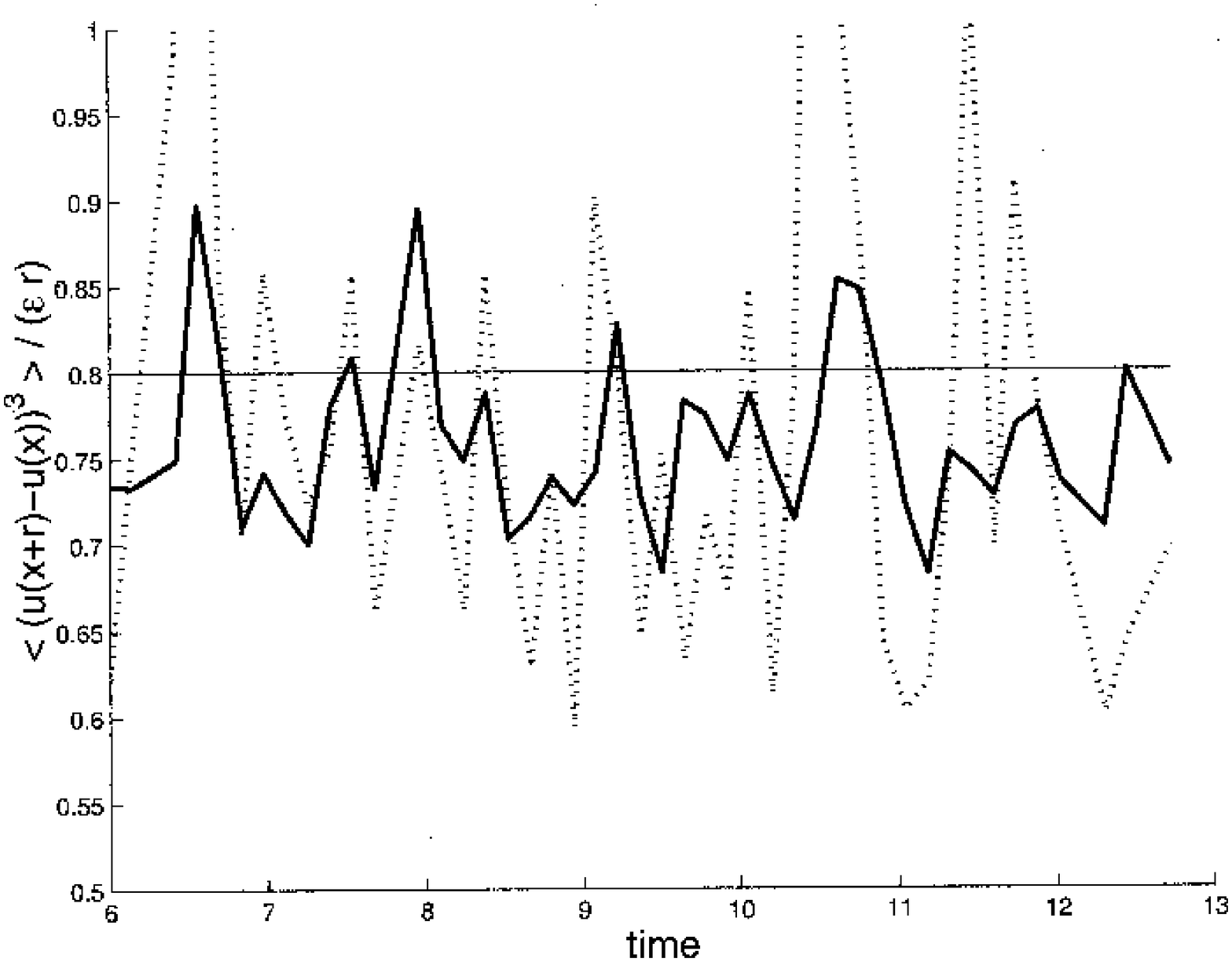}
\caption{The angle-averaged (solid line) and single-direction (dotted line) 
values of the peak of the non-dimensionalized third-order longitudinal
structure function for stochastic dataset, as a function of time.}
\label{timeseries_sto}
\end{figure}

To illustrate the variance in time of the third-order longitudinal
structure function, with and without angle-averaging, we plot the peak
value as a function of time for each dataset in Figures
\ref{timeseries} and \ref{timeseries_sto}.  The solid line is the
angle-averaged value, and the dashed line is the value from a single
coordinate direction.  The angle-averaged value has a significantly
reduced variance as compared to the single direction value, but one
can see that there still is some variance from snapshot to snapshot.
Thus in order to obtained fully converged statistics, some additional
averaging is needed.  In the next section we present results combining
angle and time averaging.  

Based on the local version of the K$41^3$ laws proved in
\cite{DucRob99,Eyink02}, we expect that increasing the spatial
resolution would allow us to obtain converged statistics from a single
snapshot when used with angle-averaging.  However, we could not expect
such convergence without angle-averaging.  This is because even in an
isotropic flow, individual snapshots are not necessary isotropic -
only the ensemble of all snapshots is guaranteed to be isotropic.

We note that the stochastic dataset (Fig. \ref{timeseries_sto}) shows a 
larger variation from snapshot to snapshot when compared to the 
deterministic dataset (Fig. \ref{timeseries}). This is true for both the 
angle-averaged and single direction quantities shown in the figures, 
suggesting that the stochastic dataset produces data with a slightly 
larger variance in time, as expected.

\subsection{Time averaged results \label{tresults}}

We now look at the 4/5ths law using both angle-averaging (which
extracts the isotropic component of the statistics) and
time-averaging (to remove the variance observed from snapshot to
snapshot).  The time-average is taken from 60 snapshots taken over 6
eddy turnover times.
The results are shown in
Fig. \ref{angle-time_comp}.  The two datasets produce nearly identical
results at all scales, even though the large scale forcing is
quite different.  The peak value of the stochastic and
deterministic datasets are .755 and .752, respectively.

Thus we conclude that flows with similar geometry and Reynolds number
have the same underlying isotropic component at all scales, at least
up to third order statistics.

\section{Conclusions \label{disc}}
We have proposed a new, computationally efficient and easily implemented
means of extracting isotropic statistics from an arbitrarily forced
flow. As a first test of the method, we averaged the third-order structure 
functions over sufficiently many angles and discovered that the K$41^3$ 
relations are obtained, with tolerable variance, from a $single$ snapshot 
of homogeneous flow with either stochastic or deterministic forcing. This 
is a stronger result than was predicted by the original Kolmogorov 
ensemble approach or even the Nie-Tanveer version of \cite{NieTan99}. 
It appears that the results are, in fact, approaching the $local$ versions 
of K$41^3$ proposed in \cite{DucRob99,Eyink02}. 

\begin{figure}
\centering
\includegraphics[scale=.6]{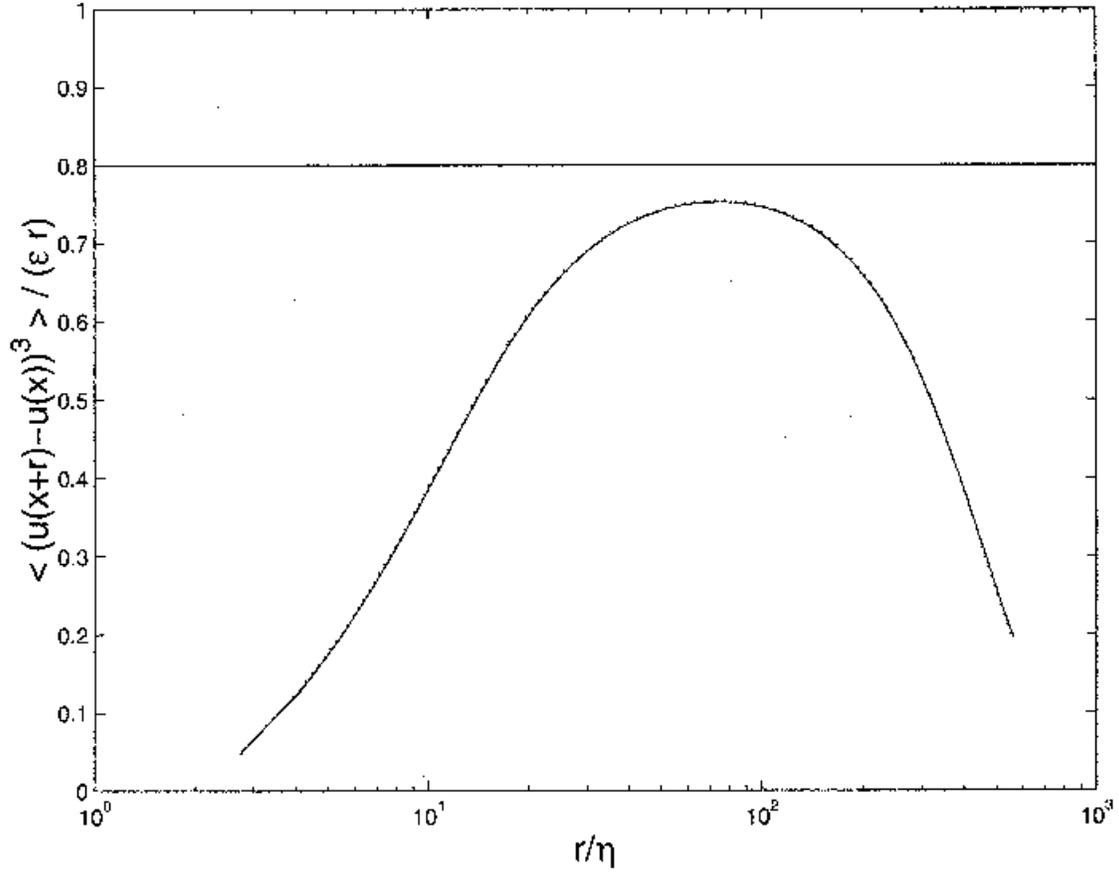}
\caption{Time and angle-averaged third-order structure function normalized by 
$\varepsilon r$, for the deterministic 
dataset (solid line) and the stochastic dataset (dotted lines).  The two 
curves are almost indistinguishable.}
\label{angle-time_comp}
\end{figure}

Using our procedure to extract the isotropic component, we are able to
separate the effect of anisotropy from the effect of finite-Reynolds
number on the statistics of the flow. This is an important point to
make in the debate on how the two effects contaminate the inertial
range. Once the anisotropy is eliminated, a more fruitful study of
finite-Reynolds number effects can be made. It is clear from
Fig. \ref{angle-time_comp} that the Reynolds numbers are still not
sufficient to give the wide inertial ranges that have been seen in
high-Reynolds number experiments.  However, it is also clear that
angle-averaging has given a significant improvement in the results.
With angle-averaging, less data is needed to obtain converged
statistics, and deterministic forcings can be used
without regard to the increased anisotropy they introduce.

The procedure we have described above can be used to investigate the
isotropic component of higher-order structure functions or any other
statistic as well. For example, the angle-averaged $n$th-order
longitudinal structure functions may be measured in this way in order
to determine scaling exponents which are truly independent of
anisotropy. This method may also be used to isolate the anisotropic
contributions themselves, as has been done in \cite{ABMP99,BLMT02}, by
subtracting from the full structure function its angle-averaged value.
Individual moments in a spherical harmonics expansion of structure
functions can be computed by introducing the basis function of interest to
the integrand in equation \ref{angave}.  In this way, the dominant
scaling in anisotropic sectors can be determined, which is important
to determine the rate of return to isotropy at small-scales. We plan
to investigate such questions in future work.

\acknowledgments
We thank Toshiyuki Gotoh for his assistance with the stochastic forcing 
procedure and for fruitful discussions.

\newpage

\newpage

\newpage

\newpage

\newpage

\newpage

\newpage

\newpage

\newpage

\newpage

\end{document}